# Vertical optical ring resonators fully integrated with nanophotonic waveguides on silicon-on-insulator substrates

Abbas Madani[1]*, Moritz Kleinert[2], David Stolarek[3], Lars Zimmermann[3], Libo Ma,[1] and Oliver G. Schmidt [1, 4]

[1]*Institute for Integrative Nanosciences, IFW Dresden, Helmholtzstr. 20, 01069 Dresden, Germany*
[2]*Fraunhofer Heinrich Hertz Institute (HHI), Einsteinufer 37, 10587 Berlin, Germany*
[3]*Leibniz-Institut fuer innovative Mikroelektronik,Technologiepark 25, 15236 Frankfurt (Oder), Germany*
[4]*Material Systems for Nanoelectronics, Chemnitz University of Technology, Reichenhainer Str. 70, 09107 Chemnitz, Germany*
*Corresponding author: a.madani@ ifw-dresden.de*

We demonstrate full integration of vertical optical ring resonators with silicon nanophotonic waveguides on silicon-on-insulator substrates to accomplish a significant step towards 3D photonic integration. The on-chip integration is realized by rolling up 2D differentially strained $TiO_2$ nanomembranes into 3D microtube cavities on a nanophotonic microchip. The integration configuration allows for out of plane optical coupling between the in-plane nanowaveguides and the vertical microtube cavities as a compact and mechanically stable optical unit, which could enable refined vertical light transfer in 3D stacks of multiple photonic layers. In this vertical transmission scheme, resonant filtering of optical signals at telecommunication wavelengths is demonstrated based on subwavelength thick walled microcavities. Moreover, an array of microtube cavities is prepared and each microtube cavity is integrated with multiple waveguides which opens up interesting perspectives towards parallel and multi-routing through a single cavity device as well as high-throughput optofluidic sensing schemes.

Optical microcavities play important roles in many diverse scientific and technological areas owing to their excellent optical properties, compactness, and potential applications [1,2]. Recently, vertically rolled-up microcavities (VRUMs), as a novel form of ring resonators [3,4], have been recognized as a promising optical unit due to the elegant fabrication procedure [5, 6], realization of new device and system concepts [7, 8], and the compatibility with on-chip integration technologies. In contrast to other types of microcavities [1], VRUMs enable the light to propagate in a plane perpendicular to the substrate surface [3,4]. The out of plane configuration makes VRUMs promising candidates for 3D photonic integration [9] where optical interconnects between multiple photonic layers are required [10]. Moreover, the hollow core of VRUMs offers an inherent channel for optofluidic applications [11,12], but has also been suggested for efficient cooling in densely packed electronic circuitry [13]. VRUMs have been investigated in fundamental studies [3,4,14,15,16], for optoelectronic applications [9,17,18,19] as well as optical sensors [11,12,20,21]. Optical coupling between waveguides and microtube cavities has been achieved by manually placing a single or two fibers in the vicinity of a tube's surface to evanescently excite the cavity modes [9,22,23]. However, this method suffers from several drawbacks such as mechanical vibrations, low reproducibility, lack of fabrication control and system instability. In another approach a single VRUM was picked from a mother substrate and directly placed onto an on-chip waveguide for optical coupling [24,25]. However, this method is also serial, time consuming and as such does not allow for any massively parallel integration scheme. In this context, it is of high interest to develop techniques to fully and monolithically integrate VRUMs with optical waveguides in order to realize on-chip optical coupling, which would be a crucial ingredient for advanced photonic applications such as unification and 3D stacked chip integration. The well-controlled fabrication process including the ability to precisely tune the distance between the VRUMs and waveguides might be considered as an indispensable asset over currently available serial and manually controlled VRUM integration schemes [12,22-25].

In this work, we experimentally demonstrate fully integrated VRUMs with Si nanowaveguides on silicon-on-insulator substrates. The integrated system is compact and mechanically stable which enables out of plane coupling between the waveguides and VRUMs and allows for 3D applications in silicon photonics. Optical characterization of these resonators reveals an extinction ratio of the integrated VRUM as high as 19 dB. Furthermore, we demonstrate the integration of VRUMs on multiple waveguides allowing for parallel and multi-routing through a single microcavity on chip.

The on-chip Si nanowaveguides were fabricated from a 220 nm thick silicon layer on 2 µm thick $SiO_2$, using 248 nm deep ultra violet (DUV) lithography and reactive ion etching (RIE) techniques [26]. Each nanowaveguide has two grating couplers at both ends to send light into the nanowaveguide and collect the transmission signal. The realization of fully integrated VRUMs relies on previously reported fabrication techniques [23,27] and are prepared by rolling up 2D differentially strained $TiO_2$ nanomembranes into 3D microtubes on the Si nanowaveguides. The integration of VRUMs on nanowaveguides is schematically outlined in Fig. 1 (a)-(d).

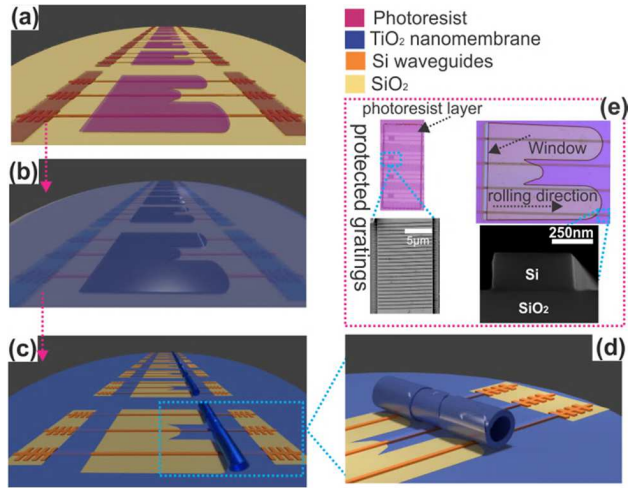

Fig.1. Schematic process of fully integrated VRUMs with optical waveguides. (a) Patterned photoresist arrays on waveguides and grating couplers are prepared by photolithography. (b) Differentially strained $TiO_2$ is deposited onto the waveguide chip. (c) The $TiO_2$ nanomembrane rolls up into microtubular structures after dissolution of the photoresist patterns. (d) Magnification of a single integrated VRUM on top of waveguides. (e) (left-panel) optical microscopy and SEM images of the photoresist-protected grating coupler; (right-panel) microscopic image of angle-deposited $TiO_2$ nanomembrane on a photoresist pattern, having an open window for photoresist removal and SEM of the cross-section of a silicon nanophotonic waveguide.

In brief, 1 μm thick photoresist (AR-P 3510, Allresist GmbH) as a sacrificial layer was spin-coated onto the nanophotonic chip surface. Standard optical photolithography was then employed to fabricate an array of U-shaped photoresist patterns for VRUMs preparation [16] and an array of rectangular-shaped photoresist patterns to protect the grating couplers, as shown in Fig. 1(a). The cross-section of a single mode Si nanowaveguide and a protected grating coupler are shown in the bottom panels of Fig.1 (e). Afterwards, a strained $TiO_2$ nanomembrane (115 nm thick) was deposited by electron beam evaporation, as shown in Fig.1 (b). $TiO_2$ was deposited by evaporating Ti metal in an oxygen atmosphere. The deposition was carried out at different rates to generate a differential strain in the nanomembrane [23,27,28]. First, a thickness of 15 nm was deposited at a low rate (0.3 nm/s), and subsequently a thickness of 100 nm was deposited at a high rate (3.8 nm/s). A 75° glancing angle was used during the deposition resulting in an uncovered small window at the left edge of the patterned photoresist [27], as is shown in the right top panel of Fig.1 (e). The photoresist was then dissolved by an organic solvent (Dimethyl sulfoxide), causing the left-to-right roll-up of the $TiO_2$ nanomembrane to release the built-in strain (Fig.1(c-d)). These integrated $TiO_2$ VRUMs possess many intriguing features such as a high refractive index (n~2.2), transparency in the visible and IR spectral ranges as well as bio- and CMOS-compatibility [23]. The rolled-up VRUMs can be separated into three segments: a bridge-like segment which is elevated above the substrate and connects two thicker "arms" (consisting of more tube windings) that remain in contact with the substrate (Fig.1(c-d)). The gap between the VRUM and the waveguides can be tuned by changing the thickness of the photoresist and the number of turns performed by the rolled-up microtube.

The fabricated VRUMs were characterized by optical and scanning electron microscopy (SEM), as displayed in Fig. 2. The top view optical microscopy image shows a rolled-up 3D microtube integrated on multiple nanowaveguides (Fig. 2(a)). The top and side view SEM images in Fig. 2(b-c) reveal detailed information about the tube windings and the gap between the tube and the waveguide. Figure 2(b) highlights the high quality of the fabrication process. The nanomembrane is rolled-up tightly, so that adjacent windings are in close contact with each other. The middle segment is elevated and bridges over the nanowaveguide (Fig.2. (c)) thus ensuring efficient light coupling from the nanowaveguide to the VRUM and vice versa. The prepared VRUM has a diameter of ~16 μm, a length of ~180 μm, and four windings in the "arms", resulting in a wall thickness of about 460 nm. The tightly rolled nanomembrane results in a compact microtube wall excellently suited to carry optical ring resonator modes. The height of the waveguide is ~220 nm and the thickness of the photoresist, i.e. the distance between the tube bottom and the substrate, is around ~1 μm. Thus, the gap between the tube bottom and the waveguide is expected to be around ~780 nm. Due to the thin tube wall and the pronounced evanescent field provided by the nanowaveguides, this gap allows for effective light coupling between the tube and nanowaveguide [29]. Multiple microtube cavities integrated on waveguide arrays (WgAs) are shown in Fig. 2(d), and one long rolled-up microtube sitting on five WgAs (consisting of four closely spaced waveguides (Fig. 2 (g)) and three single waveguides is magnified in Fig. 2(e). This demonstrates the feasibility to integrate a large number of VRUMs on multiple waveguides and WgAs at arbitrary positions. In principle, each VRUM could be integrated on equally spaced waveguides, and light in each waveguide can couple to a ring resonator mode within the VRUM (see Fig.2 (f)) and detected by a detector array for e.g. massively parallel optofluidic detection [30].

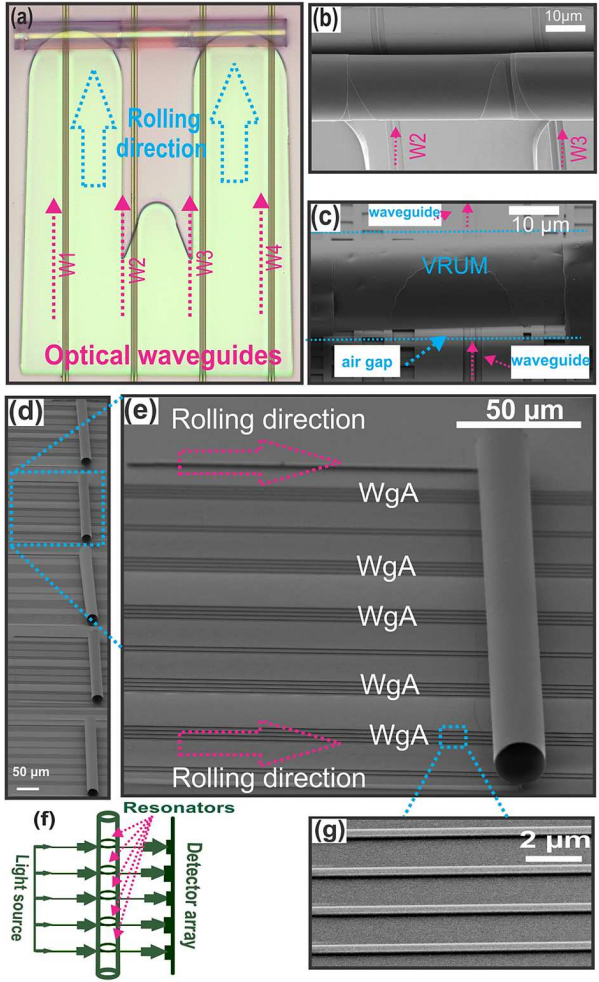

Fig. 2. (a) Top view optical microscopy image of a VRUM fully integrated with nanowaveguides on a single chip. (b) Magnification of the center bridge-like part of the microtube (top view) revealing tight and compact windings. (c) Close-up side-view shows an air gap between VRUM and waveguide as a result of the rolled-up U-shaped pattern on the photoresist. (d) SEM image of several fully integrated long VRUMs spanning over waveguide arrays (WgA). (e) Magnification of single integrated VRUM bridging across several WgAs and single waveguides. (f) Conceptual illustration of a ring resonator array along a VRUM. (g) Magnification of one part of a WgA, consisting of four closely spaced waveguides.

Optical characterization was performed in the telecom relevant spectral range. The measurement configuration is shown in Fig.3 (a). It is based on a vertical *fiber -in* and *fiber- out* setup, where two cleaved SMF-28 fibers were vertically placed to couple light in and out through a grating coupler with an incident angle of ~13° . A tunable infrared laser was used as the optical source, and a point detector recorded the intensity variation of the transmission signal. The polarization of the light to be coupled into the waveguide was set by a polarization controller (PC) which was positioned after the tunable laser. By adjusting the orientation of the linear polarization of the incoming IR laser light, the coupling between the waveguide and resonator was optimized. Successful coupling to ring resonator modes was manifested by sharp intensity drops in the transmission spectrum through the waveguide. Figure 3 (b) shows three measured transmission spectra of optical waveguides coupled at different positions of a VRUM. In this measurement, the incoming light was polarized parallel to the VRUM axis. The two transmission spectra for W1 and W3 originate from waveguides sitting below the two thick "arms" of the VRUM. Five fundamental modes are visible with mode numbers ranging from $m$ = 64 to 68 as identified by simulations previously described in Ref. [31]. These two transmission spectra exhibit very similar resonant modes due to the same diameter and wall thickness of the VRUM in these two places. The major optical modes appear at ~1501.6 nm, ~1522.9 nm, ~1544.3 nm, ~1565.8 nm, and ~1587.5 nm with an extinction ratio (ER) as high as 19 dB at 1565.8 nm. As illustrated in Fig. 3 (b), the ERs for the W1 and W3 cases are approximately the same because of the equal gap between the VRUM and W1 and W3. By assuming a 2D ring geometry with a diameter D of ~16 µm (derived from the SEM image), and a refractive index n ~2.2, the free spectral range (FSR), i.e. the distance between two adjacent fundamental peaks, is determined by the simple equation $FSR = \frac{\lambda_{res}^2}{\pi nD}$, where $\lambda_{res}$ is the wavelength of the resonant mode. The FSR is calculated to be ~21.5 nm at 1544 nm which is in excellent agreement with the experimental value (~ 21nm).

In addition, the transmission spectrum including five resonant modes, was also recorded for the nanowaveguide W2 sitting under the bridge-like segment of the VRUM. At this position the VRUM has thinner walls with smaller outer diameter leading to slightly shorter wavelengths than for the W1 and W3 cases. Figures 1 to 3 emphasize one of the key features of VRUMs: their multiplexing ability. A single long $TiO_2$ VRUM can be used as several vertical ring resonators independently at several positions along the tube axis, which is impossible to achieve with any other type of on-chip cavity such as planar microring, microdisk, or microsphere cavities. Additional transmission measurements were carried out for other nanowaveguide-VRUM systems exhibiting similar results to those presented in Fig. 3 (b) with mode positions and FSRs persistently agreeing with calculations. This validates the feasibility of monolithic mass production of VRUMs. In our work, we used $TiO_2$ as an example material but the fabrication process can be easily adapted to VRUMs made out of other materials such as Si, $SiO_X$, $SiN_x$, $Y_2O_3$, etc.

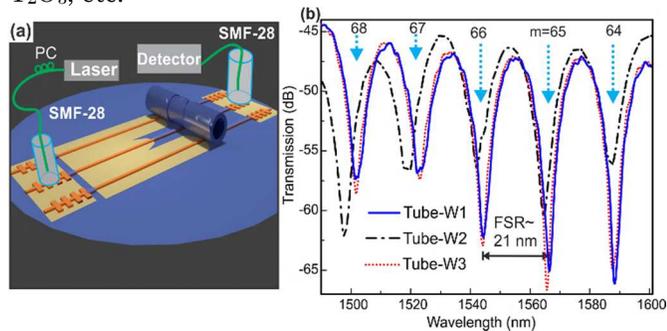

Fig.3. (a) Schematic of the *fiber-in and fiber-out* measurement set-up. (b) Transmission spectra measured through different waveguides coupled with the same VRUM at different positions W1 to W3 as defined in Fig. 2(a). Calculated mode positions for *m*= 64 to 68 are marked by dotted blue arrows.

In conclusion, we have experimentally demonstrated monolithically integrated VRUMs on top of nanowaveguides to accomplish a first important step towards 3D photonic integration. Optical characterization of these ring resonators reveals extinction ratios of up to 19 dB. Our work demonstrates an intriguing vertical coupling scheme to achieve resonant filtering of optical signals at telecom wavelengths. Moreover, we have shown that fully integrated rolled-up microtubes span across many nanowaveguides and function as vertical ring resonators at several positions along the tube axis, thus serving as a novel platform for multiplexing, optofluidic bio-sensing and opto-mechanical applications.

The authors thank R. Engelhard, S.Harazim C. Kupka, D. Karnaushenko, E. S. Gharehnaz, and K. Jung for technical support. The authors also greatly appreciate S. Böttner and M. R. Jorgensen for helpful discussions. This work was supported by the Volkswagen Foundation (I/84072) and the DFG research group No. FOR 1713.